\documentclass[%
twocolumn,
amsmath,amssymb,
aps,
prstab,
floatfix,
nofootinbib,
raggedbottom,
raggedfooter,
nobalancelastpage
]{revtex4-1}

\usepackage{times}
\usepackage{tikz,pgfplots,textcomp,booktabs,float}
\usepackage[absolute,overlay]{textpos}

\newcommand{\app}{\mathrm{\sim}}
\newcommand{\ut}[1]{\,\mathrm{#1}}

\newlength\figureheight 
\newlength\figurewidth 
\usetikzlibrary{plotmarks}

\begin{document}

\title{Luminosity Scans for Beam Diagnostics}

\author{M. Hostettler}
 \email{michi.hostettler@cern.ch}
\affiliation{%
Albert Einstein Center for Fundamental Physics, University of Bern, Switzerland\\
and CERN, CH-1211 Geneva 23, Switzerland
}%
\author{K. Fuchsberger}
\author{G. Papotti}
\author{Y. Papaphilippou}
\affiliation{%
CERN, CH-1211 Geneva 23, Switzerland
}%
\author{T. Pieloni}
\affiliation{%
Particle Accelerator Physics Laboratory, Institute of Physics, EPFL, Lausanne, Switzerland
}%

\begin{abstract}
A new type of fast luminosity separation scans (``Emittance Scans'') was introduced at the CERN Large Hadron Collider (LHC) in 2015. The scans were performed systematically in every fill with full-intensity beams in physics production conditions at the Interaction Point (IP) of the Compact Muon Solenoid (CMS) experiment. They provide both transverse emittance and closed orbit measurements at a bunch-by-bunch level. The precise measurement of beam-beam closed orbit differences allowed a direct, quantitative observation of long-range beam-beam PACMAN effects, which agrees well with numerical simulations from an improved version of the TRAIN code.
\end{abstract}

\maketitle

\section{Introduction}
The transverse beam emittance, and hence the transverse beam size, is a key parameter defining the luminosity in a collider. For high-intensity beams at the highest energies, it is a quantity challenging to measure with beam instrumentation devices in a non destructive way, as the physical beam sizes are very small due to adiabatic shrinking, and the stored beam energy prohibits the use of invasive devices like wire scanners.

Complementary to beam instrumentation, the convoluted beam width at an interaction point, where the two beams collide, can be measured by separating the beams and observing the change in luminosity. This technique was pioneered by Simon van der Meer \cite{vdm} at the CERN Intersecting Storage Rings, and is commonly used for calibrating the experiment's luminosity monitors. However these ``Van der Meer scans'' are done in dedicated machine cycles with special conditions (isolated bunches, no crossing angle) for maximum precision on the absolute luminosity calibration. Such dedicated calibration cycles are typically done once or twice per year of running, and they take several hours of machine time per experiment; for monitoring the beam conditions in regular operation, their overhead is too large.

In the following, we present a new approach to luminosity scans, called ``Emittance Scans'', which can be performed in regular physics production conditions within a few minutes. Analysis of the bunch-by-bunch luminosity measurements during these scans enables deriving transverse emittances and closed orbit offsets at a bunch-by-bunch level\footnote{Though not the primary merit (and not further expanded on in this article), a ``Van der Meer scan'' like analysis for the absolute luminosity calibrations is also possible on Emittance Scans, and is potentially useful for tracking the long-term stability of the calibration from one dedicated calibration cycle to the next one.}.

Emittance Scans were first commissioned at the LHC in 2015 and have been done regularly ever since. As they depend only on the linearity of the luminosity measurements and the transfer function of the used steering magnets, they are fully independent of any beam instrumentation commonly used for transverse emittance measurements. Therefore, the scans provide important complementary data, at an accuracy compatible with LHC requirements ($\app 10\ut{\%}$ on bunch-by-bunch transverse emittance \cite{designreport}). Accurate and independent transverse emittance measurements were crucial for following up on the efforts to maximize the luminosity yield in the past years of LHC operation. Examples include the quantification of the transverse emittance optimizations in the LHC injector complex in 2015, or the understanding of the observed asymmetry in delivered luminosities between the two high-luminosity experiments ATLAS and CMS in 2016.

In this article, we present the methodology and results from Emittance Scans performed in 2015-2017 LHC operation. In Sec.~\ref{sec:bbintro}, we give an overview of the operational setup at the LHC and the resulting beam-beam effects. The methodology and analysis of emittance scans is described in Secs.~\ref{sec:emitgauss}-\ref{sec:emitscanerr}. Finally, we present the measured effective emittances at the LHC in Sec.~\ref{sec:emitresult} and compare them to complementary measurements from beam instrumentation. The measured closed orbit separations at the interaction points are shown in Sec.~\ref{sec:bborbit} and compared to simulations; there, we also discuss the recent improvements to the TRAIN simulation code to accurately predict bunch-by-bunch beam-beam effects at the LHC.

\section{Bunch-by-Bunch diagnostics at the LHC}\label{sec:bbintro}
\subsection{Luminosity Measurements}
In the following, the main observable used for the analysis of Emittance Scans is the bunch-by-bunch luminosity as measured by the LHC experiments. Due to technical limitations, a bunch-by-bunch luminosity measurement at a high rate ($0.5\ut{Hz}$) was only available from the CMS experiment; therefore, the emittance scans were performed at this experiment.

The absolute accuracy of the CMS online luminosity measurements is typically 5-10\%, with an linearity better than 2.5\% for the luminosity range of a typical emittance scan after correction \cite{cmslumi}. The online luminosity data used for this study is provided courtesy of the CMS collaboration.

\subsection{Sources of Bunch-by-Bunch Differences}
For physics production at the LHC, each of the two rings is typically filled with more than $2000$ bunches \cite{designreport}. The bunches are mostly spaced by $25\ut{ns}$, but longer gaps separate them in ``trains'' of e.g. 96 or 144 bunches to allow for injection and extraction kicker rise times both in the injectors and the LHC. Certain effects and systems affect the bunches differently depending on their position in the train. For example, electron-cloud effects generally degrade the beam parameters towards the end of the train. Also, as explained in the following, the number of parasitic long-range beam-beam encounters depends on the position in the train. Hence, bunch-by-bunch beam diagnostics are crucial to optimize the performance of the LHC.

\subsection{Beam-Beam Effects}
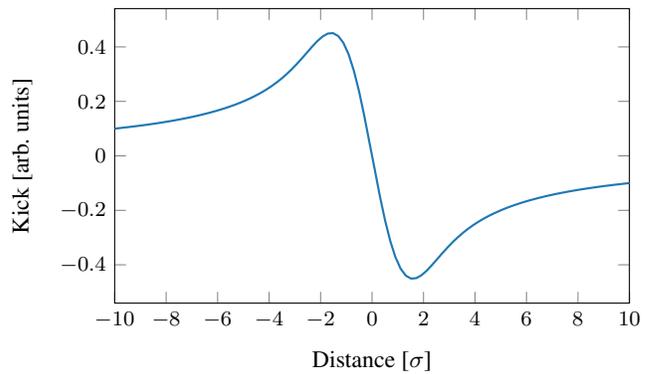
\begin{figure}[h]
\centering
\definecolor{color0}{rgb}{0.12156862745098,0.466666666666667,0.705882352941177}
\begin{tikzpicture}
\begin{axis}[small,
enlarge x limits=false,
width=0.47\textwidth,
height=5.5cm,
xlabel={Distance [$\sigma$]},
ylabel={Kick [arb.~units]},
]
\addplot[color0, thick, domain=-10:10, samples=100]{-(1/x)*(1-exp(-x*x/2))};
\end{axis}
\end{tikzpicture}
\caption{The beam-beam force for round beams. The amplitude is in arbitrary units, the separation in units of the r.m.s. beam size.}
\label{fig:beambeam}
\end{figure}

When the two beams share a common beam pipe around the 4 LHC Interaction Points (IPs), parasitic collisions outside of these points are avoided by crossing the beams at an angle. The plane in which this crossing angle is applied is referred to as the ``crossing plane'' (vertical for ATLAS/IP1, horizontal for CMS/IP5), while the orthogonal plane is referred to as the ``separation plane''. Even though direct parasitic collisions are avoided by the crossing angle, the bunches still are affected by the fields of the other beam. The functional shape of this beam-beam force is shown in Fig.~\ref{fig:beambeam}. Next to the IP, the long-range parasitic encounters typically happen at separations of $8$-$10\sigma$.

At the LHC, there are a maximum number of 45 long-range encounters over a length of $\app 168\ut{m}$ at either side of an IP until the two beams are fully separated in individual beam pipes again. In the first $\app 60\ut{m}$ around the IPs there are 15 long-range encounters in the drift space and the inner triplets, with a beam separation given by the crossing angle of the closed orbit (vertical in ATLAS, horizontal in CMS). Thereafter, the design orbits of the two beams are separated horizontally by the separation and the recombination dipoles over $\app 108\ut{m}$ by $19.4\ut{cm}$ \cite{designreport}.


Due to the gaps between the bunch trains in the LHC filling schemes \cite{beambeam}, bunches at the beginning and at the end of each bunch train are missing some long-range beam-beam encounters. This leads to a different net kick on these bunches, and hence, to bunch-to-bunch differences in orbit, tunes, and chromaticities. These so-called ``PACMAN'' effects have been predicted first in design studies of the SSC \cite{pacmanssc,pacmanssc2}, and have since been observed in the Tevatron \cite{pacmantev} and in the LHC \cite{beambeam,pacman,pacman2,tatianapacman}. Also, due to the asymmetric position of the ALICE and LHCb experiments in the LHC, some bunches do not collide at all in these IPs (``SuperPACMAN'' bunches).

While not hindering LHC operation, these PACMAN effects can affect the beam life time, and lead to bunch-to-bunch differences in orbit, tune and chromaticity, which can not be mitigated by global correctors (which do not act on individual bunches). The differences in orbit lead to bunch-to-bunch differences in the separation of the two beams at the IP, which can be measured through emittance scans. Despite of the small impact on the overall luminosity of these separations, they are a good benchmark for the overall long-range beam-beam effects and PACMAN patterns in the machine.

\section{Emittance Scans for Gaussian Bunches}\label{sec:emitgauss}
For Gaussian beams, in the presence of a beam separation, the luminosity of a colliding bunch pair is given by\footnote{This calculation neglects the ``hourglass effect'' \cite{lumi}, and is thus only valid for short bunches ($\sigma_z \ll \beta^*$). For the LHC beam parameters considered, this effect is at the level of 1\%.} \cite{lumi}
\begin{equation}\label{eqn:lumi}
{\cal L} = \frac{f_\text{rev} N_1 N_2}{2 \pi \Sigma_x \Sigma_y}\, {\cal S}
\end{equation}
where $N_{1,2}$ are the bunch intensities, $f_\text{rev}$ is the revolution frequency, and $\Sigma_x$, $\Sigma_y$ are the convoluted beam sizes in the $x$, $y$ plane (including the effect of the crossing angle in the crossing plane). 

In the presence of a separation $d$, the \emph{separation factor} ${\cal S}$ is given by
\begin{equation}\label{eqn:sep}
{\cal S} = \exp\left(\frac{-d^2}{2 \Sigma_{d}^2}\right)
\end{equation}
where $\Sigma_d$ is the convoluted beam size in the plane in which the separation $d$ is applied.

\begin{figure}[h]
\centering
\setlength{\figurewidth}{5.5cm}
\setlength{\figureheight}{3.3cm}
%
%
\begin{tikzpicture}

\begin{axis}[%
small,
width=0.951\figurewidth,
height=3.5cm,
at={(0\figurewidth,0\figureheight)},
y tick label style={/pgf/number format/.cd,%
          scaled y ticks = false,
          set thousands separator={},
          fixed},
      x tick label style={/pgf/number format/.cd,%
          scaled x ticks = false,
          fixed},
xlabel near ticks,
ylabel near ticks,
scale only axis,
xmin=0,
xmax=180,
xlabel={Time [s]},
ymin=0,
ymax=3,
max space between ticks=30pt,
try min ticks=5,
xminorticks=true,
yminorticks=true,
ylabel={Luminosity [Hz/$\mu$b]},
axis background/.style={fill=white},
legend style={at={(1,1)},anchor=north east,legend cell align=left,align=left,draw=white!15!black},
title style={font=\small},xlabel style={font={\small}},ylabel style={font=\small},legend style={font=\small},ticklabel style={font=\small}
]
\addplot [color=black,solid]
  table[row sep=crcr, y expr=\thisrowno{1}/2200]{%
0	4647.50109240001\\
1.826	4645.59528710001\\
2.852	4645.8372356\\
4.688	4646.08086310001\\
5.516	4648.3858904\\
7.546	4646.74621940002\\
8.372	4646.0280582\\
10.398	4646.90919099999\\
11.226	4646.72580309999\\
13.252	4645.65129700001\\
15.092	4647.1817364\\
15.934	4645.8200253\\
17.99	4647.162342\\
18.816	4644.9225259\\
20.65	4647.7259282\\
21.682	4645.552965\\
23.51	4646.9717003\\
24.536	4644.9316659\\
26.366	4644.8127202\\
27.266	4646.51053160001\\
29.138	4644.4317447\\
30.174	4503.8691512\\
32.002	3635.71752850001\\
34.028	2166.546637\\
34.86	1105.36313221\\
36.686	677.60634313\\
37.716	614.47925013\\
39.54	615.90082478\\
40.566	615.569357969999\\
42.416	616.928629099999\\
43.45	617.754794030001\\
45.276	617.548067310001\\
47.102	618.06083167\\
48.13	619.023424149999\\
49.962	619.03124276\\
50.988	660.469555260001\\
52.816	954.72182132\\
53.854	1450.28555567\\
55.68	1717.10001767\\
56.506	1723.69433444\\
58.532	1722.17995534\\
59.378	1723.30594761\\
61.424	1722.95171454\\
62.254	1728.85452694\\
64.076	1728.57835831\\
66.114	1730.60397015\\
66.96	1729.17618542\\
68.802	1752.16917152\\
69.834	2165.39734383\\
71.682	2986.39078151\\
72.724	3546.04834739999\\
74.552	3604.3844143\\
75.376	3602.9037007\\
77.402	3603.631356\\
78.242	3603.3660497\\
80.27	3603.8081466\\
82.1	3602.87184420001\\
83.124	3601.9656122\\
84.954	3605.2054878\\
85.778	3605.3776183\\
87.806	3839.1673517\\
88.63	4387.7491572\\
90.672	4625.18158479999\\
91.496	4640.42764090001\\
93.33	4639.8431068\\
94.36	4640.95767559999\\
96.184	4641.21404920001\\
97.212	4639.8414672\\
99.04	4640.96821260001\\
101.068	4639.69339910001\\
101.902	4640.69597170001\\
103.728	4638.7921907\\
104.756	4632.2107209\\
106.584	4440.6162301\\
107.614	3894.69559970001\\
109.44	3550.74190749999\\
110.466	3540.94459469999\\
112.294	3542.38201580001\\
113.126	3543.09467710001\\
115.162	3542.9304953\\
116.992	3546.60753840001\\
118.018	3548.3501599\\
119.844	3555.9854772\\
120.87	3555.6001437\\
122.696	3539.1207882\\
123.524	3106.34154361001\\
125.554	2274.36498375\\
126.42	1742.14414227\\
128.45	1679.85220489\\
129.276	1676.92197421\\
131.104	1677.78603965\\
132.136	1681.28374527\\
133.964	1682.74851245\\
135.994	1684.96429079\\
136.826	1686.12785162\\
138.662	1689.43560507\\
139.714	1693.09549359\\
141.61	1525.11625388\\
142.496	1035.72585764\\
144.324	676.102788790001\\
145.354	592.197062119999\\
147.19	594.085448899999\\
148.216	595.025390490001\\
150.044	595.285201619999\\
152.08	594.455971090001\\
152.984	594.672484270001\\
154.9	596.849242959998\\
155.762	597.258826489999\\
157.626	597.878348240001\\
158.51	616.740675350001\\
160.392	888.81577328\\
161.47	1711.65284167\\
163.294	3130.49896111\\
164.162	4290.33293410001\\
166.04	4622.3080053\\
168.114	4636.96270639999\\
168.948	4635.78614229999\\
170.794	4635.4672672\\
171.854	4634.98369580002\\
173.716	4636.50949759999\\
174.572	4633.0570756\\
176.424	4634.9912097\\
177.486	4634.4783435\\
179.318	4635.81698679999\\
180.346	4637.5897741\\
182.174	4637.6186335\\ 
};
 
\end{axis}
\end{tikzpicture}%
\caption{Luminosity evolution during a scan in one plane.}
\label{fig:exlumi}
\end{figure}
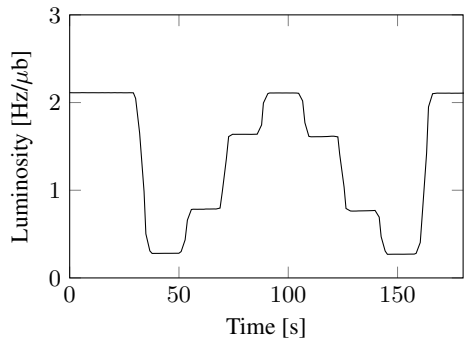

\begin{figure}[h]
\centering
\setlength{\figurewidth}{5.5cm}
\setlength{\figureheight}{3.3cm}
\input{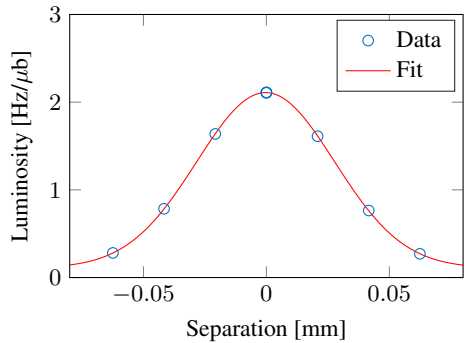}
\caption{Fitted beam profile from a scan in one plane.}
\label{fig:exfit}
\end{figure}

By scanning the separation $d$ in each plane and fitting a Gaussian of the form
\begin{equation}\label{eqn:fit}
{\cal L}(d_{u}) = {\cal L}_\text{peak}\,\exp\left(\frac{-(d_{u}-d_{u0})^2}{2 \Sigma_{u}^2}\right)\;\;\;\text{for}\;\;\;u=x,y 
\end{equation}
to the measured luminosity, both the convoluted beam sizes $\Sigma_{x,y}$ and the initial (parasitic) separations $d_{x0,y0}$, can be determined. An example luminosity measurement during a scan with 7 separation steps is shown in Fig.~\ref{fig:exlumi}, and the resulting fit is given in Fig.~\ref{fig:exfit}. If a bunch-by-bunch luminosity measurement is available, the fitting can be done independently for each colliding bunch pair.

To derive the transverse emittances $\varepsilon_{x,y}$ from the convoluted beam sizes $\Sigma_{x,y}$, it is assumed that the beam sizes of Beam 1 and Beam 2 are equal\footnote{The sizes of the two LHC beams are typically within $5-10\ut{\%}$ considering independent measurements, e.g. from the synchrotron light monitors. If the emittances of the two beams are different, this analysis yields the \emph{effective} emittance: the average of the two emittances of a colliding bunch pair.}. In this case, the transverse beam size $\sigma_\text{sep}$ in the separation plane is given by
\begin{equation}\label{eqn:capsigma_sep}
\Sigma_\text{sep} = \sqrt{2} \sigma_\text{sep}
\end{equation}

In the crossing plane, the longitudinal bunch profile is projected onto the convoluted beam size $\Sigma_\text{xing}$ through the crossing angle. For a Gaussian longitudinal profile with bunch length $\sigma_z$ and half crossing angle $\alpha$, the convoluted beam size is given by
\begin{equation}\label{eqn:capsigma_xing}
\Sigma_\text{xing} = \sqrt{2 \sigma_\text{xing}^2 \cos^2(\alpha) + 2 \sigma_{z}^2 \sin^2(\alpha)}
\end{equation}

The effective normalized transverse emittances $\varepsilon_{x,y}$ are then derived from the transverse beam sizes $\sigma_{x,y}$ accounting that the dispersion as the IPs is negligible\footnote{The nominal dispersion at the LHC IPs is 0. After optics corrections, the dispersion beating is less than $1.2 \cdot 10^{-2}\ut{m}$ at the IPs \cite{tobiasoptics}. This changes the beam size by less than 1\%.} and $\beta^*=\beta^*_x=\beta^*_y$:
\begin{equation}\label{eqn:emit_sigma}
\varepsilon_{u} = \frac{\gamma}{\beta^{*}} \sigma_{u}^2\;\;\;\text{for}\;\;\;u=x,y 
\end{equation}
where $\gamma$ is the relativistic gamma factor and $\beta^*$ is the value of the $\beta$-function at the IP.

\section{Emittance Scan Parameters}\label{sec:emitscanparam}
The scan parameters, namely the separation steps to scan and the time to acquire data at each step, were optimized to provide an adequate measurement while keeping the scans as short as possible. The steps were chosen to be equidistant and symmetric with respect to the initial position of the beams. A total effective separation of $\app 3 \sigma$ is aimed for, corresponding to a separation factor of $S\approx 0.1$ and hence a change in luminosity by a factor of $\app 10$. Half of this total separation is applied symmetrically to the orbit of both beams with an opposite sign. Due to the crossing angle, $\Sigma_\text{xing}$ is larger and the effect of the separation is smaller in the crossing plane. This can be accounted for by applying a larger separation in the crossing plane.

The operational scan parameters established for the LHC based on these constraints are given in Table \ref{tbl:params}. The scan range is given in units of the ``LHC nominal'' beam size $\sigma_\text{nom}$ (assuming a nominal emittance of $\varepsilon_\text{nom} = 3.5 \ut{\mu m}$), while the actual beam emittance was $\varepsilon \approx 2.5 \ut{\mu m}$ as of 2016 \cite{evian16}. Scans were done regularly at the CMS experiment as the experiment could deliver bunch-by-bunch luminosity readings at a high rate. Considering the integration time and the time required to displace the beams, a pair of scans can be done in less than 4 minutes.

\begin{table}[hbt]
   \centering
   \caption{Scan Parameters}
    \begin{ruledtabular}
   \begin{tabular}{lc}
          Number of separation steps         & $7$        \\
           Integration time per step       & $10\ut{s}$        \\
           Beam separation (crossing plane)        & $2.5 \ut{\sigma_\text{nom}}$        \\
           Beam separation (separation plane)       & $3.5 \ut{\sigma_\text{nom}}$        \\
   \end{tabular}
   \end{ruledtabular}
   \label{tbl:params}
\end{table}

\section{Emittance Scans for Longitudinally Non-Gaussian Bunches}\label{sec:emitnongaussian}
\begin{figure}[h]
\centering
\setlength{\figurewidth}{5.5cm}
\setlength{\figureheight}{2.7cm}
\input{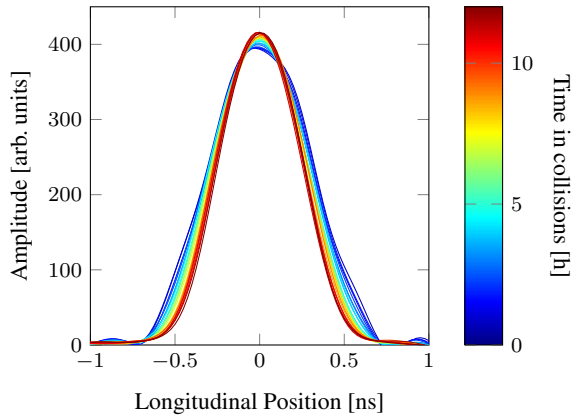}
\caption{Longitudinal profiles in LHC fill 4964 for beam 1, bunch 142 over $12\ut{h}$ in collisions, measured by a Wall Current Monitor pick-up sampled at 40 GS/s. Other bunches (including those in beam 2) behave similarly.}
\label{fig:exprofile}
\end{figure}

In the nominal LHC cycle, the bunch length is increased artificially during the energy ramp by injecting RF phase noise to guarantee beam stability \cite{rfnoise}. This changes the longitudinal bunch profile from a Gaussian to a distribution with a larger core and lighter tails, which can be represented as a q-Gaussian with $q \approx 0.85$ \cite{stefania}. Over the course of several hours in collisions, the longitudinal distribution resumes a more Gaussian shape (Fig.~\ref{fig:exprofile}).

If the longitudinal distribution is not Gaussian, the simple factorization of the convoluted beam size in Eqn.~\ref{eqn:capsigma_xing} is no longer valid for the crossing plane \cite{georgebb}. While still assuming a Gaussian distribution in the transverse dimensions, the luminosity during a scan of the separation $d$ in the crossing plane fulfills

\begin{align}\label{eqn:comp_long}
{\cal L} &\propto \exp\left(\frac{-d^2}{4 \sigma_\text{xing}^2}\right) \cdot {\cal C}(d)\\
{\cal C}(d) &= \int\limits_{-\infty}^{\infty}\, \mathrm{d}s\, (f_1 * f_2)(2s)\,\exp\left(\frac{-s \sin(\alpha) (s\,\sin(\alpha)-d)}{\sigma_\text{xing}^2}\right)\label{eqn:comp_c}
\end{align}
where $\alpha$ is the half crossing angle, $f_{1,2}$ are the longitudinal distribution functions for beam 1 and beam 2 respectively, and $*$ denotes a convolution.

It is worth noting that $\alpha = 0$ yields ${\cal C}(d) = \mathrm{const.}$ and the separation dependency reduces to the \emph{separation factor} as shown in Eqs.~\ref{eqn:sep}, \ref{eqn:capsigma_sep}. This also proves that the longitudinal distribution only affects scans in the crossing plane.

If the longitudinal distribution is measured for both beams, the convolution and the integration in Eq.~\ref{eqn:comp_c} can be done numerically, and a non-linear regression of Eq.~\ref{eqn:comp_long} to the measured scan points yields $\sigma_\text{xing}$.

\subsection{Simulations}
To benchmark this approach, emittance scans were simulated for bunches with a transverse Gaussian and longitudinal q-Gaussian distribution with $q=0.85$. This longitudinal distribution is typical for bunches in the LHC at the start of collisions.

\begin{figure}[h]
\centering
\setlength{\figurewidth}{5.5cm}
\setlength{\figureheight}{2.7cm}
\input{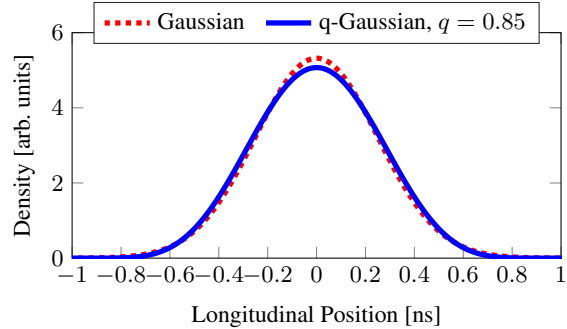}
\caption{Longitudinal q-Gaussian profile used in the simulations, compared to a Gaussian profile of the same r.m.s. width.}
\label{fig:simprofile}
\end{figure}

The simulations were done for two different sets of machine parameters and emittances, corresponding to the configurations used during the LHC 2015 and 2016 run periods. The simulated data was then analyzed using the method proposed above, and compared to the Gaussian approximation based on full-width half-maximum (FWHM) and true r.m.s. bunch length measurements. The operational bunch length measurement in the LHC is based on the FWHM method.

Results are compiled in Table \ref{tbl:simprof}. While commissioning emittance scans during 2015 LHC operation, the Gaussian approach using the operational FWHM bunch length measurement was used and a systematic error of up to 15\% in the crossing plane due to the non-Gaussian longitudinal distribution was assumed \cite{paper2015}.

For measuring low-emittance beams at a large crossing angle, as at the LHC in 2016 and 2017, using measured longitudinal bunch profiles or a true r.m.s.\ bunch length measurement becomes crucial. For the machine parameters considered, the error on the emittance measurement using a Gaussian fit with a true r.m.s.\ bunch length measurement is below $5\ut{\%}$.

\begin{table}[p]
   \centering
   \caption{Reconstruction of emittances from simulated emittance scans in the crossing plane with a longitudinal \mbox{q-Gaussian} ($q=0.85$) distribution for the LHC machine parameters used in 2015 and in 2016.}
    \begin{ruledtabular}
    \begin{tabular}{p{3cm}cc}
        \multicolumn{3}{c}{\textbf{2015 LHC machine parameters}}\\
        \multicolumn{3}{c}{$\beta^*=45\ut{cm}$, $\alpha=\pm 145\ut{\mu rad}$, $\sigma_s=7.5\ut{cm}$}\\
       \colrule
        Method & \multicolumn{2}{c}{Reconstructed Emittances} \\
        ~ & {$\varepsilon_{ref}=2\ut{\mu m}$}   & {$\varepsilon_{ref}=3.5\ut{\mu m}$} \\
       \colrule
          Gaussian fit, FWHM bunch length & $1.85\ut{\mu m}$ &  $3.34\ut{\mu m}$ \\
          Gaussian fit, r.m.s. bunch length & $2.02\ut{\mu m}$ &  $3.51\ut{\mu m}$ \\
          Direct fit using longitudinal profile & $2.00\ut{\mu m}$ &  $3.50\ut{\mu m}$ \\
   \end{tabular}
   \end{ruledtabular}
   \begin{ruledtabular}
   \begin{tabular}{p{3cm}cc}
        \multicolumn{3}{c}{\textbf{2016 LHC machine parameters}}\\
        \multicolumn{3}{c}{$\beta^*=40\ut{cm}$, $\alpha=\pm 185\ut{\mu rad}$, $\sigma_s=7.5\ut{cm}$}\\
       \colrule
        Method & \multicolumn{2}{c}{Reconstructed Emittances} \\
        ~ & {$\varepsilon_{ref}=2\ut{\mu m}$}   & {$\varepsilon_{ref}=3.5\ut{\mu m}$} \\
       \colrule
          Gaussian fit, FWHM bunch length & $1.55\ut{\mu m}$ &  $3.02\ut{\mu m}$ \\
          Gaussian fit, r.m.s. bunch length & $2.09\ut{\mu m}$ &  $3.57\ut{\mu m}$ \\
          Direct fit using longitudinal profile & $2.00\ut{\mu m}$ &  $3.50\ut{\mu m}$ \\
   \end{tabular}
   \end{ruledtabular}
   \label{tbl:simprof}
\end{table}

\subsection{LHC Machine Studies}
The impact of a changing longitudinal distribution on emittance scan was directly probed in 2016 during the commissioning of the longitudinal bunch flattening at the LHC \cite{juanthesis}.

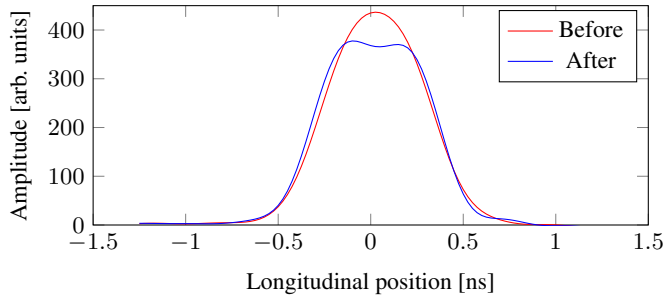
\begin{figure}[p]
\centering
%
%
\definecolor{mycolor1}{rgb}{0.12156862745098,0.466666666666667,0.705882352941177}
\definecolor{mycolor3}{rgb}{1,0.498039215686275,0.0549019607843137}
\definecolor{mycolor4}{rgb}{0.172549019607843,0.627450980392157,0.172549019607843}
\definecolor{mycolor2}{rgb}{0.83921568627451,0.152941176470588,0.156862745098039}
\definecolor{mycolor5}{rgb}{0.580392156862745,0.403921568627451,0.741176470588235}

\begin{tikzpicture}

\begin{axis}[%
small,
width=0.5\textwidth,
height=4.5cm,
xlabel={Longitudinal position [ns]},
ylabel={Amplitude [arb.~units]},
ylabel near ticks,
max space between ticks=30pt,
try min ticks=5,
at={(0\textwidth,0cm)},
xmin=-1.5,
xmax=1.5,
ymin=0,
ymax=450,
axis background/.style={fill=white},
title style={font=\small},xlabel style={font={\small}},ylabel style={font=\small},legend style={font=\small},ticklabel style={font=\small}
]
\addplot [color=red,solid]
  table[row sep=crcr]{%
-1.25	2.8010432119063\\
-1.225	3.24855467759828\\
-1.2	3.54866833063306\\
-1.175	3.67375684298625\\
-1.15	3.62521454567715\\
-1.125	3.43511530410463\\
-1.1	3.15332018361914\\
-1.075	2.84888157493275\\
-1.05	2.58542496063308\\
-1.025	2.424950648512\\
-1	2.39871557566359\\
-0.975	2.51727332465349\\
-0.95	2.76619188628977\\
-0.925	3.10985219819114\\
-0.9	3.49550710338922\\
-0.875	3.87042910676898\\
-0.85	4.19049926753391\\
-0.825	4.42958663916119\\
-0.8	4.58618220242825\\
-0.775	4.68981036163675\\
-0.75	4.80044233728221\\
-0.725	5.00394899856254\\
-0.7	5.41510715574156\\
-0.675	6.16562048977542\\
-0.65	7.41719530665245\\
-0.625	9.34938425856374\\
-0.6	12.1653669480448\\
-0.575	16.1066957852918\\
-0.55	21.4316071910773\\
-0.525	28.4280599213089\\
-0.5	37.3980781858938\\
-0.475	48.6335955620087\\
-0.45	62.389433887321\\
-0.425	78.8628747064798\\
-0.4	98.1391731613683\\
-0.375	120.178250211645\\
-0.35	144.777513231073\\
-0.325	171.566969142151\\
-0.3	200.012534323097\\
-0.275	229.447178521743\\
-0.25	259.102097037428\\
-0.225	288.170838685062\\
-0.2	315.862831835658\\
-0.175	341.46948559932\\
-0.15	364.414460196521\\
-0.125	384.292608643046\\
-0.1	400.875423380939\\
-0.075	414.112948208407\\
-0.05	424.100577913919\\
-0.025	431.027922125482\\
0	435.120052102774\\
0.025	436.588891843098\\
0.05	435.583824931208\\
0.075	432.152636875447\\
0.1	426.246762977927\\
0.125	417.729157879275\\
0.15	406.416470106385\\
0.175	392.128110782433\\
0.2	374.749995202916\\
0.225	354.294927949787\\
0.25	330.935019514034\\
0.275	305.03358538488\\
0.3	277.142639137916\\
0.325	247.972288335051\\
0.35	218.333836844361\\
0.375	189.084255232684\\
0.4	161.046554794204\\
0.425	134.936924065793\\
0.45	111.314351900486\\
0.475	90.5417622832617\\
0.5	72.7674151186204\\
0.525	57.9439060739487\\
0.55	45.8606777651215\\
0.575	36.1839049786491\\
0.6	28.5199009266346\\
0.625	22.4639606760183\\
0.65	17.6412583739631\\
0.675	13.7479323734839\\
0.7	10.5584089786193\\
0.725	7.92229078227533\\
0.75	5.76013292500409\\
0.775	4.03294157142321\\
0.8	2.7202069226071\\
0.825	1.79674366354658\\
0.85	1.2235648001355\\
0.875	0.925157500841835\\
0.9	0.814715537540137\\
0.925	0.783043629300003\\
0.95	0.725518108149114\\
0.975	0.552342582631796\\
1	0.195764677555994\\
1.025	-0.362180065456175\\
1.05	-1.1205505731808\\
1.075	-2.03259867287162\\
1.1	-3.03859793588563\\
1.125	-4.06598073437764\\
1.15	-5.04911238980197\\
1.175	-5.92902627277163\\
1.2	-6.66991613505252\\
1.225	-7.25820401080576\\
};
\addlegendentry{\small{Before}};

\addplot [color=blue,solid]
  table[row sep=crcr]{%
-1.25	3.45049852225096\\
-1.225	3.41174390745159\\
-1.2	3.31712192572913\\
-1.175	3.18745203264667\\
-1.15	3.05329985537202\\
-1.125	2.93742149908966\\
-1.1	2.85853908985638\\
-1.075	2.82076603334546\\
-1.05	2.81600602512559\\
-1.025	2.82480414331702\\
-1	2.82366266386716\\
-0.975	2.79476101273758\\
-0.95	2.73182061996021\\
-0.925	2.64476799432784\\
-0.9	2.56248859183102\\
-0.875	2.5343648434337\\
-0.85	2.61437140557369\\
-0.825	2.85548091392991\\
-0.8	3.29631793335487\\
-0.775	3.95303249309294\\
-0.75	4.81969725875156\\
-0.725	5.87273921242965\\
-0.7	7.08933147493337\\
-0.675	8.47265648571543\\
-0.65	10.0800709979241\\
-0.625	12.0557605110285\\
-0.6	14.6549656336431\\
-0.575	18.2489051473378\\
-0.55	23.3172201520908\\
-0.525	30.4238129634274\\
-0.5	40.1555467479626\\
-0.475	53.0486626911742\\
-0.45	69.521279036315\\
-0.425	89.7867742530495\\
-0.4	113.77817885852\\
-0.375	141.119244812137\\
-0.35	171.10896489328\\
-0.325	202.750090609911\\
-0.3	234.822462003562\\
-0.275	265.982371618002\\
-0.25	294.896333655592\\
-0.225	320.372912983142\\
-0.2	341.485481701274\\
-0.175	357.677309453109\\
-0.15	368.821172218414\\
-0.125	375.223880023901\\
-0.1	377.581686278568\\
-0.075	376.892818447621\\
-0.05	374.314662600297\\
-0.025	371.021621002775\\
0	368.036632145917\\
0.025	366.103859679938\\
0.05	365.583575039419\\
0.075	366.404656130304\\
0.1	368.071394987243\\
0.125	369.72959477288\\
0.15	370.284000032278\\
0.175	368.538815844291\\
0.2	363.361261723852\\
0.225	353.832830907428\\
0.25	339.383451504663\\
0.275	319.86492347822\\
0.3	295.590448282001\\
0.325	267.31070591451\\
0.35	236.134085536675\\
0.375	203.414731346407\\
0.4	170.609629392373\\
0.425	139.136715478764\\
0.45	110.236613301528\\
0.475	84.8659071687116\\
0.5	63.6257353268177\\
0.525	46.7430619625874\\
0.55	34.0871618467906\\
0.575	25.218013586862\\
0.6	19.4877195192948\\
0.625	16.1264747767701\\
0.65	14.342575676667\\
0.675	13.418639434583\\
0.7	12.773018421874\\
0.725	11.9971425630202\\
0.75	10.8678414062888\\
0.775	9.33139801217957\\
0.8	7.46900176194599\\
0.825	5.44525636072395\\
0.85	3.4586648471455\\
0.875	1.69158099455991\\
0.9	0.279879322672906\\
0.925	-0.714600441331734\\
0.95	-1.30327431974116\\
0.975	-1.56040381933298\\
1	-1.60892536440155\\
1.025	-1.59098914283259\\
1.05	-1.63012438922903\\
1.075	-1.83242390598139\\
1.1	-2.25176665352865\\
1.125	-2.88956172101332\\
1.15	-3.70646613407451\\
1.175	-4.62352154902647\\
1.2	-5.54769020661685\\
1.225	-6.38586413640141\\
};
\addlegendentry{\small{After}};

\end{axis}
\end{tikzpicture}%
\caption{The measured longitudinal profile before and after the flattening.}
\label{fig:profile-before-after}
\end{figure}

During these tests, a sinusoidal RF phase modulation was applied to compensate for the shrinking caused by synchrotron radiation damping by flattening the bunch profiles. The distribution was measured before and after, the change is shown in Fig.~\ref{fig:profile-before-after}.

\begin{figure}[p]
\centering
%
%
\definecolor{mycolor1}{rgb}{0.12156862745098,0.466666666666667,0.705882352941177}
\definecolor{mycolor2}{rgb}{1,0.498039215686275,0.0549019607843137}
\definecolor{mycolor4}{rgb}{0.172549019607843,0.627450980392157,0.172549019607843}
\definecolor{mycolor3}{rgb}{0.83921568627451,0.152941176470588,0.156862745098039}
\definecolor{mycolor5}{rgb}{0.580392156862745,0.403921568627451,0.741176470588235}

\begin{tikzpicture}

\begin{axis}[%
small,
width=0.5\textwidth,
height=5cm,
xlabel={Separation [mm]},
ylabel={Luminosity [Hz/$\mu$b]},
ylabel near ticks,
max space between ticks=40pt,
try min ticks=5,
at={(0\textwidth,0cm)},
/pgf/number format/1000 sep={},
/pgf/number format/fixed,
xmin=-0.1,
xmax=0.1,
ymin=0,
ymax=3.5,
axis background/.style={fill=white},
legend style={legend cell align=left,align=left,draw=white!15!black,at={(0.5,0.95)}, anchor=south,font=\small},
legend columns=2,
title style={font=\small},xlabel style={font={\small}},ylabel style={font=\small},ticklabel style={font=\small}
]
\addplot [color=red,only marks,mark=o,mark options={solid}]
  table[row sep=crcr]{%
-0.041648884	0.815277777777778\\
-0.034540736	1.23118214285714\\
-0.02743259	1.70706785714286\\
-0.020324442	2.19671111111111\\
-0.013216294	2.65979285714286\\
-0.006108148	2.96155\\
0.001	3.05031785714286\\
0.008108148	2.94868928571429\\
0.015216294	2.64025185185185\\
0.022324442	2.17027777777778\\
0.02943259	1.655325\\
0.036540736	1.17794642857143\\
0.043648884	0.76692962962963\\
};
\addlegendentry{\small{Data Before}\;\;\;};

\addplot [color=red,solid]
  table[row sep=crcr]{%
-0.1	0.00179241305756893\\
-0.095	0.00349833215184778\\
-0.09	0.00671635905364436\\
-0.085	0.0126090796280737\\
-0.08	0.0230425211843049\\
-0.075	0.0408639801928866\\
-0.07	0.0701852498728521\\
-0.065	0.116586677424359\\
-0.06	0.187110896332708\\
-0.055	0.289886515664111\\
-0.05	0.433238310731309\\
-0.045	0.624222041756796\\
-0.04	0.866673367590866\\
-0.035	1.15906344896246\\
-0.03	1.49265877004382\\
-0.025	1.85060241165667\\
-0.02	2.20847266844525\\
-0.015	2.53659233127159\\
-0.01	2.80391712317733\\
-0.005	2.98285571560965\\
0	3.05401067109328\\
0.00500000000000001	3.00970991633784\\
0.01	2.85541570716433\\
0.015	2.60863676467708\\
0.02	2.29564948457509\\
0.025	1.94689132709091\\
0.03	1.59213911720466\\
0.035	1.2565007839917\\
0.04	0.957911994364466\\
0.045	0.706361889057557\\
0.05	0.5046200297685\\
0.055	0.349937294847371\\
0.06	0.236112848371355\\
0.065	0.155422890472658\\
0.07	0.100102584246036\\
0.075	0.0632741353789156\\
0.08	0.0393691443229894\\
0.085	0.0241803362506598\\
0.09	0.0146982873419974\\
0.095	0.00886251528563512\\
0.1	0.00531066391296818\\
};
\addlegendentry{\small{Simulation Before}};

\addplot [color=blue,only marks,mark=o,mark options={solid}]
  table[row sep=crcr]{%
-0.0412341405459629	0.857125925925926\\
-0.0341259925459629	1.2533037037037\\
-0.0270178465459629	1.69955185185185\\
-0.0199096985459629	2.15321428571429\\
-0.0128015505459629	2.52086666666667\\
-0.00569340454596289	2.76472142857143\\
0.00141474345403711	2.82977037037037\\
0.00852289145403711	2.70873333333333\\
0.0156310374540371	2.38881851851852\\
0.0227391854540371	1.9766\\
0.0298473334540371	1.51741071428571\\
0.0369554794540371	1.08447142857143\\
0.0440636274540371	0.710618518518519\\
};
\addlegendentry{\small{Data After}};

\addplot [color=blue,solid]
  table[row sep=crcr]{%
-0.1	0.00156748204090332\\
-0.095	0.00326449373568761\\
-0.09	0.00652407055460619\\
-0.085	0.0125342777023122\\
-0.08	0.0231782816603381\\
-0.075	0.0412908689225359\\
-0.07	0.0709008000148148\\
-0.065	0.11736413377733\\
-0.06	0.187253906486389\\
-0.055	0.287866385790063\\
-0.05	0.426259816573664\\
-0.045	0.60785891562867\\
-0.04	0.834804304944091\\
-0.035	1.10435195459839\\
-0.03	1.40769679256832\\
-0.025	1.72959074316096\\
-0.02	2.04904237572677\\
-0.015	2.34122038152731\\
-0.01	2.58044255377935\\
-0.005	2.74384093879904\\
0	2.81501501605103\\
0.00500000000000001	2.78683565948342\\
0.01	2.66267436018923\\
0.015	2.4557326506576\\
0.02	2.18668337981099\\
0.025	1.88026249098182\\
0.03	1.56161024290215\\
0.035	1.25306966184053\\
0.04	0.971920424994804\\
0.045	0.729261680211729\\
0.05	0.530015784444268\\
0.055	0.373835337372569\\
0.06	0.256573293844402\\
0.065	0.171930800841232\\
0.07	0.112937550209574\\
0.075	0.0730364698534792\\
0.08	0.0466992478186948\\
0.085	0.029633557276013\\
0.09	0.0187147898574847\\
0.095	0.0117806485311519\\
0.1	0.00739178726083597\\
};
\addlegendentry{\small{Simulation After}};

\end{axis}
\end{tikzpicture}%
\caption{The scan data and the simulation based on the measured profiles. The effects of the bunch profile change are well reproduced. The transverse emittances of the two simulated scans are identical.}
\label{fig:scan-before-after}
\end{figure}
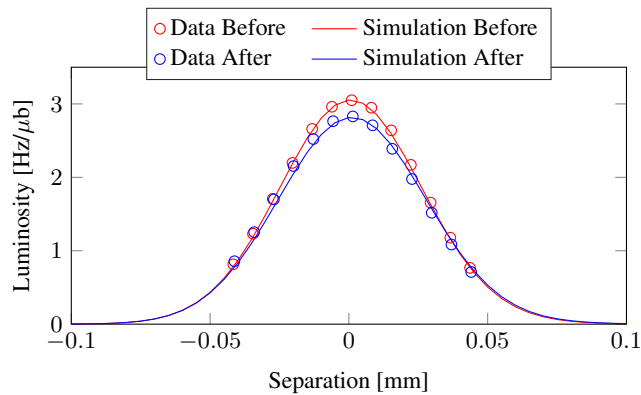

Before and after the longitudinal bunch flattening, an emittance scan was done in the crossing plane. Results are shown in Fig.~\ref{fig:scan-before-after}, and compared to the yield of Eq.~\ref{eqn:comp_long} considering the change in longitudinal distribution and intensity, but assuming no change in transverse emittance. The agreement shows that the longitudinal distribution is well accounted for.

\subsection{Retrospective Correction}
An operational measurement of the longitudinal bunch profiles was not available at the LHC as of 2017; longitudinal bunch profiles could only be logged by manual intervention on demand. However, since the machine parameters and the longitudinal distribution were reproducible, the difference between the reconstruction based on a FWHM measurement and the real emittance reduces to a constant offset at first order. This allows using the operational FWHM bunch length measurement and correcting the resulting emittance by an offset to be determined by a reference profile measurement and numerical simulations.

For emittance scans at the start of collisions in 2016, a simulation shows that the emittance in the crossing plane is underestimated by $0.5\pm0.1\ut{\mu m}$ for the operational machine parameters and a typical bunch length of $\sigma_\text{z}=7.8\pm0.5\ut{cm}$. This is a constant offset in emittance, and is only weakly dependent on the actual bunch length. Neglecting the bunch length dependence and applying a correction of $+0.5\ut{\mu m}$, the residual error on the emittance is less than $5\ut{\%}$ for the range of bunch lengths and emittances used at the LHC.


\section{Systematic Errors}\label{sec:emitscanerr}
At high luminosity, systematic errors are largely predominant, while statistical errors on the luminosity measurement are negligible. The error sources considered and their contribution to the total error of the derived emittances are given in Table \ref{tbl:errors}. For the crossing plane, it is assumed that the longitudinal bunch shape is taken into account up to a residual error of $5\ut{\%}$ as described in the previous section.

\begin{table}[hbt]
   \centering
   \caption{Errors on Emittances From Luminosity Scans}
   \begin{ruledtabular}
   \begin{tabular}{lcc}
        \textbf{Error source} & \textbf{separation}   & \textbf{crossing} \\
        \textbf{} & \textbf{plane}   & \textbf{plane} \\
       \colrule
          luminosity non-linearity \cite{cmslumi}         & 5\% &  5\%        \\
          steering magnet calibration \cite{cmslumi}     & 5\% &  5\%        \\
          $\beta^*$ (optics uncertainty) \cite{tobiasoptics}        & 3\% &  3\%         \\
          head-on beam-beam kick \cite{jorgbb,cmsbb}	& 2\% &  2\%         \\
          dynamic $\beta^*$ (beam-beam) \cite{cmsbb}         & 2\% &  2\%         \\
          transverse distribution \cite{stefania}        & 5\% & 5\%        \\
          crossing angle    & - & 5\%        \\
          longitudinal distribution        & - & 5\%        \\
       \colrule
         \textbf{combined error}         & 9.6 \% & 11.9\%        \\
   \end{tabular}
   \end{ruledtabular}
   \label{tbl:errors}
\end{table}

It is to be noted that most systematic error sources only affect the absolute scales of the derived beam sizes and emittances. Relative bunch-to-bunch differences and other fit parameters (in particular the initial separation $d_0$) are not affected. Also, only the luminosity non-linearity due to pile-up effects and the longitudinal bunch shape are expected to change over the course of a fill, and therefore to possibly affect the time evolution of the emittance.

\section{Emittance Measurements}\label{sec:emitresult}
\subsection{Convoluted Emittances}
The effective convoluted transverse emittances $\varepsilon_\text{c}=\sqrt{\varepsilon_\text{x}\varepsilon_\text{y}}$ at the start of collisions, averaged over all bunches in the LHC, are shown Figs.~\ref{fig:emit-compare-2015},\ref{fig:emit-compare-2016} for the LHC proton physics runs in 2015 and 2016 respectively. For comparison, the other emittance measurements available for high-intensity beams are also displayed: the synchrotron light monitors \cite{bsrt} and the reconstruction from the luminosities measured by the ATLAS and CMS experiments. The available measurements agree better than $20\ut{\%}$ in convoluted emittance, which is the expected level of systematic uncertainties \cite{evian16}.
\begin{figure}[h]
\centering
%
%
\definecolor{mycolor1}{rgb}{0.00000,0.44700,0.74100}%
\definecolor{mycolor2}{rgb}{0.92900,0.69400,0.12500}%
\definecolor{mycolor3}{rgb}{0.85000,0.32500,0.09800}%
\definecolor{mycolor4}{rgb}{0.49400,0.18400,0.55600}%
\definecolor{mycolor5}{rgb}{0.46600,0.67400,0.18800}%
\begin{tikzpicture}

\begin{axis}[%
small,
width=0.47\textwidth,
height=5.1cm,
unbounded coords=jump,
xmin=4300,
xmax=4600,
try min ticks=8,
/pgf/number format/1000 sep={},
max space between ticks=30pt,
ymin=1.1,
ymax=4,
xlabel={Fill Number},
ylabel={\small{Emittance [$\mu$m]}},
axis background/.style={fill=white},
legend style={legend cell align=left,align=left,draw=white!15!black},
ylabel near ticks,
legend style={at={(0.02,0.03)},anchor=south west,legend cell align=left,align=left, text width=7.6em,text height=1.5ex},
legend columns=2,
title style={font=\small},xlabel style={font={\small}},ylabel style={font=\small},legend style={font=\small},ticklabel style={font=\small}
]
\addplot [color=mycolor3,only marks,mark size=2,mark=triangle*,mark options={solid,rotate=180,fill=mycolor4,draw=mycolor4}]
  table[row sep=crcr]{%
4322	nan\\
4323	nan\\
4332	2.81150117408124\\
4337	2.36310617789575\\
4341	2.60153020174371\\
4349	2.72643146314603\\
4356	2.79878452221267\\
4363	2.89284515199099\\
4364	2.77416536066517\\
4386	2.70308801224045\\
4391	2.95600480321335\\
4420	2.82350318819828\\
4428	2.66294684645759\\
4440	2.90162968737439\\
4444	3.03301156212496\\
4455	2.90037543188264\\
4462	2.98632320684927\\
4476	3.61473016688348\\
4479	3.20220888216755\\
4518	nan\\
4519	nan\\
4522	2.95744142255388\\
4528	3.12311039316298\\
4530	3.10762331538218\\
4532	3.07719835471984\\
4538	2.94152598857404\\
4540	2.81179394871092\\
4545	2.92064737896333\\
4557	2.92022234730726\\
4560	3.17959814050579\\
4562	3.24169258138094\\
4565	3.20104823116016\\
4569	3.06558230821836\\
};
\addlegendentry{\small{Synchrotron Light}};

\addplot [color=mycolor4,only marks,mark size=1.5,mark=square*,mark options={solid,fill=blue,draw=blue}]
  table[row sep=crcr]{%
4322	2.58242847142282\\
4323	2.56520387615215\\
4332	2.69230079525036\\
4337	2.71015958833991\\
4341	2.91522380224915\\
4349	2.92089507813351\\
4356	3.01553547256145\\
4363	3.19853977314608\\
4364	2.89477127907977\\
nan	nan\\
4391	3.16319632424877\\
4420	2.83738259684248\\
4428	2.93258893759142\\
4440	2.96842418386246\\
4444	3.1977327328279\\
4455	2.88689935450881\\
4462	2.85284206789803\\
4476	3.22067274314485\\
4479	2.82071325914736\\
4518	2.76554290885387\\
4519	3.10697730712844\\
4522	3.03999872867535\\
4528	3.20585228248172\\
4530	3.11138544143979\\
4532	3.15302557002454\\
4538	3.10291866617402\\
4540	3.07009441895546\\
4545	3.17929564358818\\
4557	3.36177160050568\\
4560	3.40830721957379\\
4562	3.43823797047932\\
4565	3.45996236619755\\
4569	3.25593200811864\\
};
\addlegendentry{ATLAS abs. lumi};

\addplot [color=mycolor5,only marks,mark size=1.5,mark=square*,mark options={solid,fill=black,draw=black}]
  table[row sep=crcr]{%
4322	3.08644525475144\\
4323	2.86259839236768\\
4332	2.99862063569996\\
4337	3.02551562930495\\
4341	3.2068243306544\\
4349	3.21542141203928\\
4356	3.29082098886775\\
4363	3.38645932467716\\
4364	3.14865463657047\\
4386	2.98863833865628\\
4391	3.21973791273449\\
4420	3.04604272199837\\
4428	3.09714020246795\\
4440	3.0214984223143\\
4444	3.062274463148\\
4455	3.14226113560288\\
4462	3.00227787798516\\
4476	3.47079421663354\\
4479	3.07744028271446\\
4518	2.98294778452338\\
4519	3.32935396806815\\
4522	3.23937515077083\\
4528	3.44732301129136\\
4530	3.37060843560497\\
4532	3.36508454968853\\
4538	3.37659949323486\\
4540	3.33315173607761\\
4545	3.46691885559221\\
4557	3.19368833182509\\
4560	3.7294570323398\\
4562	3.61693438578213\\
4565	4.09079494344733\\
4569	4.64119750937105\\
};
\addlegendentry{CMS abs. lumi};

\addplot [color=mycolor2,only marks,mark size=2,mark=diamond*,mark options={solid,fill=mycolor3,draw=mycolor3}]
 plot [error bars/.cd, y dir = both, y explicit]
 table[row sep=crcr,
x expr=\thisrowno{0},  y expr=\thisrowno{1}*(0.95)
]{%
4322	2.68315212816925	0.161436285212708	0.161436285212708\\
4323	2.78366035375516	0.292226806603842	0.292226806603842\\
4332	2.8958980209925	0.241720404691649	0.241720404691649\\
4337	2.92413281916333	0.25371063547595	0.25371063547595\\
4341	3.19420159856219	0.272156095960139	0.272156095960139\\
4349	3.19576540827511	0.26880009894803	0.26880009894803\\
4356	3.25492363447875	0.304947604396963	0.304947604396963\\
4363	3.38305051384776	0.618692127148925	0.618692127148925\\
4364	3.13281948410233	0.302695048956734	0.302695048956734\\
4386	3.15998820226121	0.534960495136156	0.534960495136156\\
4391	3.32412311391016	0.275069527872752	0.275069527872752\\
4420	3.22011559178831	0.222354785243324	0.222354785243324\\
4428	3.270565865451	0.239682122672963	0.239682122672963\\
4440	3.19767892566458	0.206929209565488	0.206929209565488\\
4444	3.19039049016658	0.230607797835638	0.230607797835638\\
4455	3.27522854855729	0.244335690080263	0.244335690080263\\
4462	3.16337746024599	0.243078907404184	0.243078907404184\\
4476	3.62230885200875	0.290357110638616	0.290357110638616\\
4479	3.20560378704942	0.254832327222217	0.254832327222217\\
4518	3.08246402367479	0.202888754419654	0.202888754419654\\
4519	3.40377602347004	0.240188286040385	0.240188286040385\\
4522	3.30203520051106	0.308764431452228	0.308764431452228\\
4528	3.53240084056958	0.25213005163605	0.25213005163605\\
4530	3.41610069840526	0.261706721051839	0.261706721051839\\
4532	3.46802384072059	0.247234382341224	0.247234382341224\\
4538	3.33454223241201	0.220507476430308	0.220507476430308\\
4540	3.25947912524074	0.236944938839209	0.236944938839209\\
4545	3.41459364457451	0.225961595661767	0.225961595661767\\
4557	3.28136409899235	0.232827984826788	0.232827984826788\\
4560	3.60908627134035	0.280671038800808	0.280671038800808\\
4562	3.5704367207029	0.253470657976862	0.253470657976862\\
4565	3.75761877220023	0.285792776152761	0.285792776152761\\
4569	3.33228904799302	0.248829622745987	0.248829622745987\\
};
\addlegendentry{Emittance Scans};

\end{axis}
\end{tikzpicture}%
\caption{The average convoluted emittances at the start of collisions in 2015 proton physics operation (only fills with emittance scans are shown).}
\label{fig:emit-compare-2015}
\end{figure}
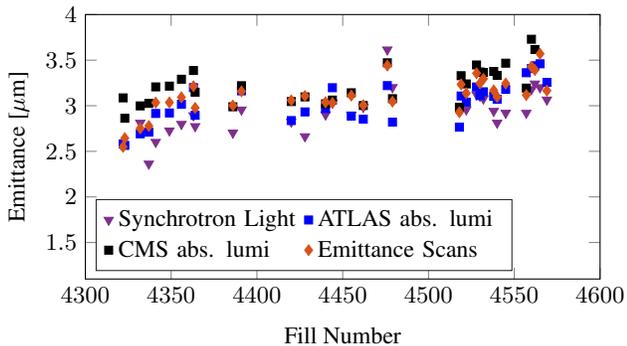
\begin{figure}[h]
\centering
\input{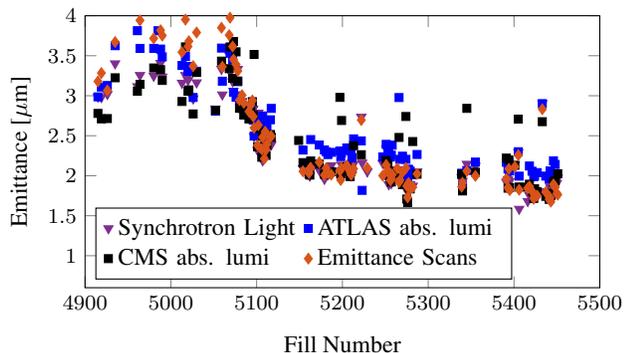}
\caption{The average convoluted emittances at the start of collisions in 2016 proton physics operation.}
\label{fig:emit-compare-2016}
\end{figure}

In 2016 as of fill 5079, the convoluted emittance was reduced from $3.4\pm0.6\ut{\mu m}$ to $2.0\pm0.3\ut{\mu m}$ at the start of collisions with the introduction of the ``Batch Compression Merging and Splitting'' \cite{bcms} beam production scheme in the injector complex. The resulting reduction in emittance is confirmed by all monitors.

\subsection{Horizontal-Vertical Emittance Asymmetry}
A difference between the ATLAS and CMS luminosities has been observed throughout 2016 LHC operation, while ideally the two experiments would receive the same luminosity from the LHC. As the beams cross in different planes in the two experiments, non-round beams can cause such an imbalance via the crossing angle (see Eqns.~\ref{eqn:capsigma_sep}, \ref{eqn:capsigma_xing}) \cite{imbalance}.

\begin{figure}[h!]
\centering
%
%
\definecolor{mycolor1}{rgb}{0.00000,0.44700,0.74100}%
\definecolor{mycolor2}{rgb}{0.85000,0.32500,0.09800}%
\begin{tikzpicture}

\begin{axis}[%
small,
width=0.47\textwidth,
height=4.45cm,
xmin=4900,
xmax=5500,
xtick={4900,5000,5100,5200,5300,5400,5500},
xminorticks=true,
try min ticks=5,
xlabel={Fill Number},
ylabel={\small{Emittance [$\mu$m]}},
ylabel near ticks,
/pgf/number format/1000 sep={},
ymin=1.1,
ymax=4.5,
yminorticks=true,
axis background/.style={fill=white},
legend style={at={(0.01,0.02)},anchor=south west,legend cell align=left,align=left,draw=white!15!black}
]
\addplot [color=mycolor1,only marks,mark=triangle*,mark options={solid,rotate=180,fill=mycolor2,draw=mycolor2}]
  table[row sep=crcr]{%
4915	3.74628221394401\\
4919	3.9783591840003\\
4926	3.07532769848425\\
4935	4.43204217225269\\
4961	5.37879049134982\\
4964	4.7721132639166\\
4980	4.27528359231921\\
4985	4.59290738775343\\
4988	4.3947257582438\\
4990	4.41948980707566\\
5013	4.0139901763084\\
5017	4.72017015036518\\
5020	3.99870084278937\\
5021	4.21156017854509\\
5026	3.98045803591851\\
5030	4.45527549724464\\
5052	5.25153038867838\\
5059	4.61004841639615\\
5060	3.97516420347951\\
5068	4.58822009507715\\
5069	4.71076272141157\\
5072	4.29279901267247\\
5073	3.97862071963623\\
5076	3.82956854660273\\
5078	3.73429748693284\\
5080	3.44015386596322\\
5083	3.50396993010031\\
5085	3.40035364797955\\
5091	3.24733313351832\\
5093	3.3682990329186\\
5095	3.48038569588602\\
5096	3.15860655156392\\
5097	3.2709540270342\\
5101	2.82845799729674\\
5102	3.11511942568872\\
5106	2.7936860887451\\
5107	3.16323811055371\\
5109	2.700300323355\\
5110	3.27364665309681\\
5111	2.78305170155484\\
5112	2.83918155652177\\
5116	3.04616265939302\\
5117	3.05428589403683\\
5149	3.14879091556991\\
5154	2.46977907938429\\
5161	2.58165757725528\\
5162	2.62823936534191\\
5163	2.57289202415304\\
5173	2.72587496633871\\
5179	2.55497177031587\\
5181	2.67543913125884\\
5183	2.5967675266875\\
5187	2.39921747045585\\
5197	2.61238819724272\\
5198	2.47663728802188\\
5199	2.6005517173789\\
5205	2.67099648944383\\
5206	2.67919251755283\\
5209	2.62096373638944\\
5210	2.53190668174094\\
5211	2.58710432355583\\
5213	2.58783901987826\\
5222	3.22104047895989\\
5223	3.01965031573909\\
5229	2.63989544356607\\
5247	2.48418990378291\\
5251	2.54873899802907\\
5253	2.46925096807791\\
5254	2.59294109223458\\
5256	2.37554586104268\\
5257	2.44104394226185\\
5258	2.47245864233167\\
5264	2.48345014444205\\
5266	2.30583509737296\\
5267	2.35283291272976\\
5270	2.55381263127656\\
5274	2.46218201122198\\
5275	2.34030772908536\\
5276	2.09727947201414\\
5277	2.35698630495572\\
5279	2.28536379339525\\
5282	2.31016721697394\\
5287	2.53320275516548\\
5339	2.19566983376308\\
5340	2.31869297410151\\
5345	2.42012837693129\\
5355	2.33165751794301\\
5391	2.37957040505613\\
5393	2.11160698753175\\
5394	2.57516075272369\\
5395	2.17919514791628\\
5401	2.44887738061082\\
5405	2.67716252475982\\
5406	2.20613324300638\\
5416	2.1685776580533\\
5418	2.2102814955943\\
5421	2.14097622607578\\
5423	2.01550147418811\\
5424	2.05992914873322\\
5427	2.05080998472252\\
5433	3.08396551207876\\
5437	2.1172670029696\\
5441	1.990461746655\\
5442	1.98563758443516\\
5443	1.92715955928609\\
5446	2.17340557500507\\
5448	2.13251008010346\\
5450	2.08195191007261\\
5451	2.04293652683965\\
};
\addlegendentry{\small{horizontal}};

\addplot [color=mycolor2,only marks,mark=triangle*,mark options={solid,fill=mycolor1,draw=mycolor1}]
  table[row sep=crcr]{%
4915	3.11403308816412\\
4919	3.10178698439124\\
4926	3.65342211316927\\
4935	3.42954584035607\\
4961	4.4007058708119\\
4964	3.63887970302442\\
4980	3.66052998126619\\
4985	3.96411209399052\\
4988	3.74541836784455\\
4990	3.59385657744147\\
5013	3.57715622174978\\
5017	3.70190099956568\\
5020	3.73813674026949\\
5021	3.65345346625138\\
5026	3.26382071972713\\
5030	3.63443800518659\\
5052	4.05626740819784\\
5059	3.60707722880035\\
5060	3.25476468844419\\
5068	3.45621307915455\\
5069	3.7539734749658\\
5072	3.44685509686701\\
5073	3.42980929075621\\
5076	3.45539254795951\\
5078	3.38902111384371\\
5080	2.9503391709745\\
5083	3.01138936283727\\
5085	2.9249004761826\\
5091	2.86038770991318\\
5093	2.89788439143011\\
5095	2.90357951168177\\
5096	2.82478147182179\\
5097	2.43889884884448\\
5101	2.43247827492135\\
5102	2.64945460579319\\
5106	2.36766853773514\\
5107	2.3110557727256\\
5109	2.22561927698359\\
5110	2.3462701724823\\
5111	2.37022914372115\\
5112	2.45334383525598\\
5116	2.42972275827536\\
5117	2.4372384861194\\
5149	nan\\
5154	2.14468908070925\\
5161	1.98693378071244\\
5162	2.04637993438882\\
5163	2.1275681940297\\
5173	2.11253622214087\\
5179	1.96504589712061\\
5181	2.01488705591538\\
5183	1.97123795858496\\
5187	2.13543147142535\\
5197	2.01465072801521\\
5198	1.91834441646083\\
5199	2.09272025107552\\
5205	2.08774007590284\\
5206	2.02885111616217\\
5209	1.97952504032282\\
5210	1.99551756185645\\
5211	1.96771751639837\\
5213	2.08714012700831\\
5222	2.67263749999489\\
5223	nan\\
5229	2.05567657218986\\
5247	2.01847239104094\\
5251	2.22629076616367\\
5253	2.1743857809162\\
5254	1.9834898091232\\
5256	2.0095419231725\\
5257	1.98050086772657\\
5258	2.05834968278663\\
5264	2.20789995634105\\
5266	2.04550112975099\\
5267	2.04166152262222\\
5270	2.03263362414187\\
5274	2.23845396015268\\
5275	1.90595736156165\\
5276	1.87323789746897\\
5277	1.96275046532894\\
5279	1.88633195240762\\
5282	1.90545875083342\\
5287	2.01517341201495\\
5339	2.04740295630439\\
5340	2.01291102795145\\
5345	2.20782469731005\\
5355	2.17987672427439\\
5391	2.0453454539302\\
5393	2.12482413408234\\
5394	2.09805869098313\\
5395	1.97330371366543\\
5401	2.28963854758092\\
5405	2.34027374436047\\
5406	1.95235715522145\\
5416	2.03795918497755\\
5418	2.03470873315622\\
5421	1.98804201646255\\
5423	1.94313412181487\\
5424	1.94052800104491\\
5427	1.97889924162673\\
5433	3.10433266464429\\
5437	1.99319467017803\\
5441	1.95393383379205\\
5442	2.00639267299\\
5443	1.97061078308028\\
5446	2.13858325459711\\
5448	2.10900758249838\\
5450	1.99851417474373\\
5451	2.01431027074128\\
};
\addlegendentry{\small{vertical}};

\end{axis}
\end{tikzpicture}%
\caption{The average horizontal and vertical emittances at the start of collisions in 2016 proton physics operation.}
\label{fig:emit-x-y}
\end{figure}
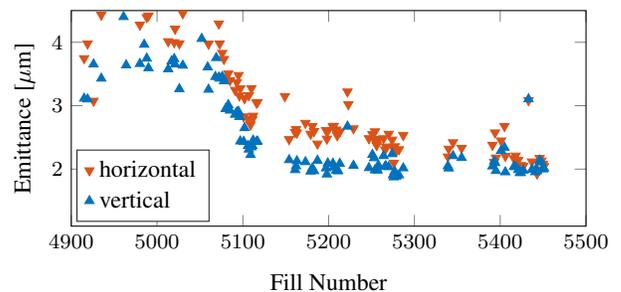

Non-round beams were indeed observed in emittance scans in 2016, as shown in Fig.~\ref{fig:emit-x-y}. The larger emittance in the horizontal plane penalizes the ATLAS experiment (crossing vertically) less than the CMS experiment (crossing horizontally). The level of the measured emittance asymmetry is consistent with the ratio of the ATLAS and CMS luminosities \cite{imbalance}.

\subsection{Bunch-by-Bunch Emittances}
Effective bunch-by-bunch emittances for a typical LHC fill are shown for all bunches in Fig.~\ref{fig:emit-bbb} and details for 6 typical bunch trains in Fig.~\ref{fig:emit-bbb-zoom}. An increase in emittance over the individual trains by $\app 0.5 \ut{\mu m}$ is visible, attributed to electron cloud effects. Also, a decreasing trend in emittance is visible over the full ring. Bunches early in the ring are injected first, and thus circulate for a longer time in the LHC at injection energy where the degradation due to intra-beam scattering is strongest \cite{fanouriamodel}.

\begin{figure}[p]
\centering
\input{bbb_emit_h.tikz}
\input{bbb_emit_v.tikz}
\caption{The bunch-by-bunch emittances of all 2556 bunches in the LHC (fill 6054).}
\label{fig:emit-bbb}
\centering
\input{bbb_emit_h_zoom.tikz}
\input{bbb_emit_v_zoom.tikz}
\caption{Zoom on Fig.~\ref{fig:emit-bbb}. The shaded error region for the emittance scan data includes both systematic and statistical uncertainties.}
\label{fig:emit-bbb-zoom}
\input{bbb_emit_lumi_zoom.tikz}
\caption{Bunch-by-bunch convoluted emittances for 6 typical bunch trains (LHC fill 6054). The shaded error region for the emittance scan data includes both systematic and statistical uncertainties.}
\label{fig:emit-bbb-lumi}
\end{figure}

The measurements from the synchrotron light monitors and the emittance scans agree well within the expected systematic and statistical uncertainties. As expanded in section \ref{sec:emitscanerr}, the systematic uncertainties affecting the absolute emittance scale for all bunches are strongly dominant over per-bunch statistical or systematic errors.

In Fig.~\ref{fig:emit-bbb-lumi}, the measured convoluted bunch-by-bunch emittances are compared to the convoluted emittances reconstructed from the bunch-by-bunch luminosity measurements in the ATLAS and CMS experiments. Again, an agreement well within the expected statistical and systematic uncertainties is observed.

\section{Closed Orbit Separation Measurements}\label{sec:bborbit}
\begin{figure}[h]
\centering
%
%
\begin{tikzpicture}

\begin{axis}[%
small,
width=0.56\textwidth,
height=3cm,
at={(0\textwidth,0cm)},
axis on top,
xmin=0.5,
xmax=3564.5,
ymin=0.5,
ymax=1.5,
ytick=\empty,
xlabel={25ns Bunch Slot Number},
axis background/.style={fill=white},
/pgf/number format/1000 sep={}
]
\addplot [forget plot] graphics [xmin=0.5,xmax=3564.5,ymin=0.5,ymax=1.5] {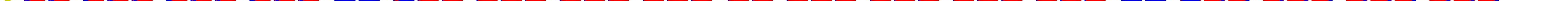};
\end{axis}
\end{tikzpicture}%
\caption{The 2556 bunch filling scheme used for LHC beam 1 in 2017 proton physics operation. Apart from the 12 non-colliding bunches (yellow), the filling scheme for beam 2 is identical. Bunches in blue are not colliding in the ALICE or LHCb experiment.}
\label{fig:fillingscheme}
\end{figure}

\begin{figure}[h]
\centering
\input{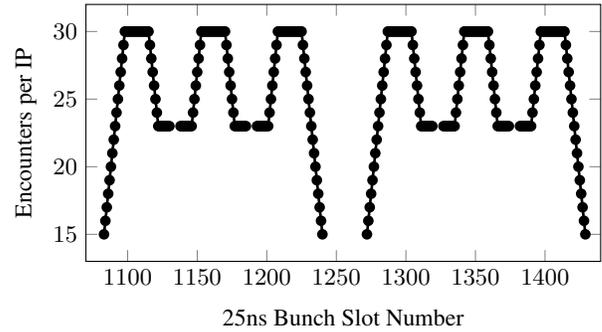}
\caption{Number of long-range beam-beam encounters in the drift space and inner triplets around ATLAS and CMS for two trains of the 2556 bunch filling scheme used in 2017 LHC operation. The pattern is identical for all trains.}
\label{fig:lrbb}
\end{figure}

For 2017 proton physics operation, PS batches of 48 bunches were joined to form trains of 96 bunches (2 PS batches) or 144 bunches (3 PS batches). The LHC filling scheme consisted of 15 trains with 144 bunches, interleaved with 4 trains of 96 bunches. Together with 12 non-colliding ``witness bunches'', this totaled to 2556 bunches per LHC beam (Fig.~\ref{fig:fillingscheme}). The number of long-range beam-beam encounters per bunch in the drift space around the ATLAS and CMS experiments for this filling scheme is shown in Fig.~\ref{fig:lrbb}. Due to PACMAN effects, the different number of beam-beam long-range encounters is expected to yield a difference in closed orbits, which can be observed using emittance scans.

\subsection{Simulations}
The TRAIN code \cite{train,arekthesis} can simulate self-consistent closed orbits on a bunch-by-bunch level under the presence of long-range beam-beam effects. These simulations take into account the filling scheme, and optionally measured intensities and emittances. The simulation gives the bunch-by-bunch closed orbit for either beam, as well as bunch-by-bunch tunes and chromaticities. The difference between the closed orbits of the two beams at the IP yields to a separation, which can be observed during emittance scans. This effect has previously also been observed at the LHC during van-der-Meer scans \cite{vdmbb}.

\begin{figure}[h]
\centering
\input{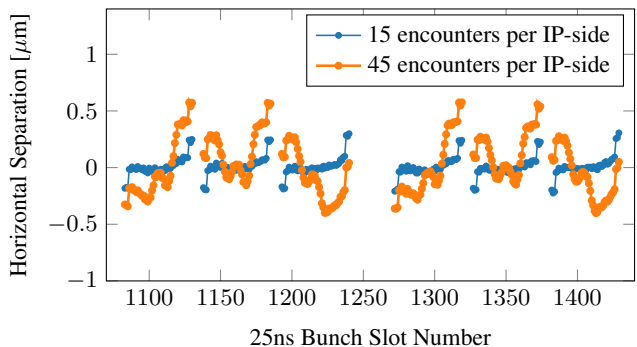}
\caption{TRAIN simulation of the closed-orbit separations in LHC fill 5976 (ATS optics, $\beta^*=40\ut{cm}$ in ATLAS and CMS, $\varepsilon=2.5\ut{\mu m}$, $N=1.15 \cdot 10^{11}\ut{ppb}$). The plot shows the expected separation in horizontal plane of CMS, where the additional long-range beam-beam encounters beyond the 15 in the inner triplets and drift space have the strongest impact.}
\label{fig:15lrbb}
\end{figure}

In earlier versions, the TRAIN simulation only took into account the 15 long-range encounters in the IP drift space and inner triplets \cite{arekthesis}. This approximation was justified for the LHC nominal design optics, but for the ``achromatic telescopic squeeze'' (ATS, \cite{ats}) optics used in the recent years of LHC operation, the effect of the long-range encounters beyond the 15th, while small, could sum up to non-negligible orbit kicks. This led to a discrepancy between measurements and simulation results in earlier studies, which were most pronounced in the horizontal (crossing) plane at CMS \cite{arekipac}. For this study, we therefore extended the TRAIN code to support an arbitrary number of long-range beam-beam encounters, taking into account both the closed and the design orbit. As shown in Fig.~\ref{fig:15lrbb}, considering the full 45 long-range encounters approximately doubles the PACMAN orbit effects in the horizontal plane at CMS, and introduces an additional fine structure over the bunch trains. A convergence study considering different numbers of long-range beam-beam encounters showed that at least 35-40 encounters per IP-side are needed to accurately reproduce this structure.

\subsection{Measurements}
\begin{figure}[p]
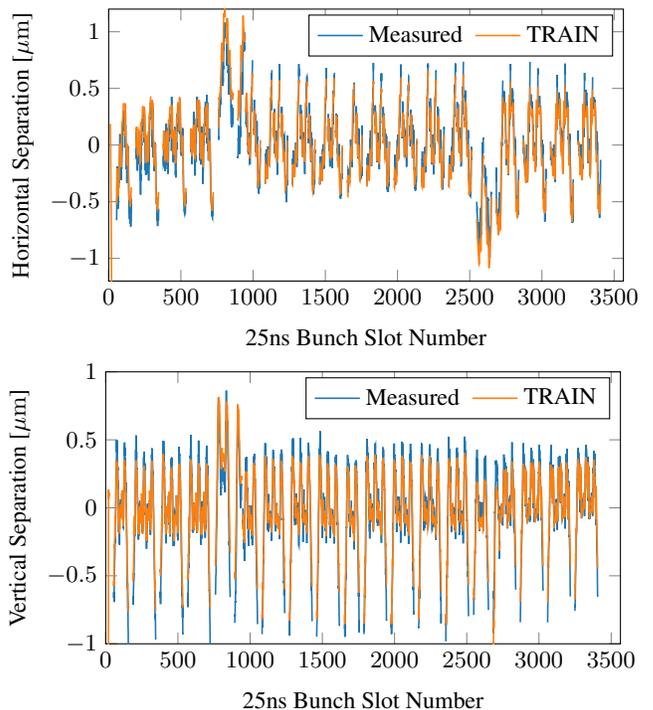

\centering
\input{sep-train-h.tikz}
\input{sep-train-v.tikz}
\caption{The bunch-by-bunch orbit separations at the CMS experiment for all bunches colliding there (fill 5976). Note that the trains missing collisions in ALICE and LHCb have different separations (in particular in the horizontal plane).}
\label{fig:sep-train}
\end{figure}

\begin{figure}[p]
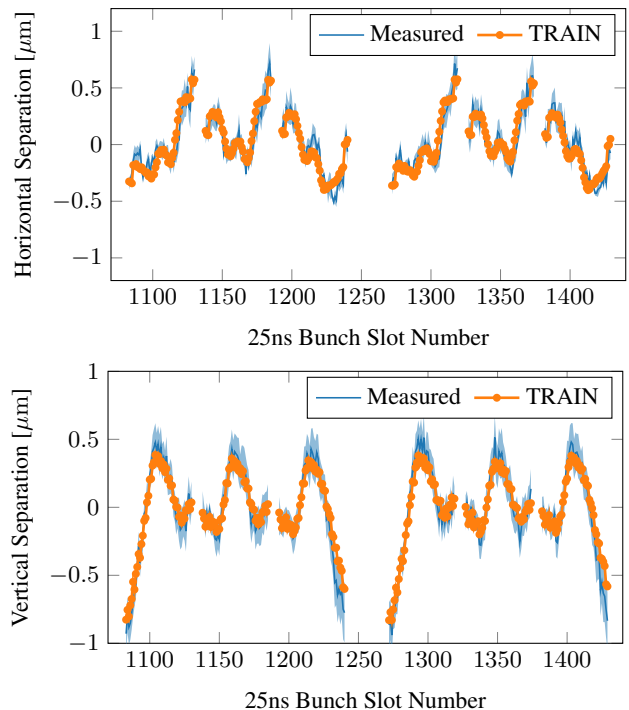

\centering
\input{sep-train-h-zoom.tikz}
\input{sep-train-v-zoom.tikz}
\caption{Zoom on Fig.~\ref{fig:sep-train}, showing two typical bunch trains colliding in all four IPs. The shapes predicted by the TRAIN simulation match the data within uncertainties.}
\label{fig:sep-train-zoom}
\end{figure}

Figs.~\ref{fig:sep-train}, \ref{fig:sep-train-zoom} show the bunch-by-bunch parasitic separations (Eqn.~\ref{eqn:fit}, $d_{x0,y0}$) measured by an emittance scan, compared to the simulation results of the TRAIN code. The simulation takes into account the measured emittances and bunch intensities.

The PACMAN bunches in the beginning and in the end of each 144 bunch train have different closed orbits due to the different number of beam-beam long-range encounters. The structure in the horizontal plane is predominately generated by long-range encounters in CMS (horizontal crossing), while the vertical structure are due to long-range encounters in ATLAS (vertical crossing).

Also, the SuperPACMAN trains missing head-on collisions in ALICE or LHCb are offset in closed orbit with respect to the trains colliding in all experiments. This is due to the use of luminosity levelling by separation in these experiments: the partially separated beams cause a coherent beam-beam kick, which the SuperPACMAN bunches miss at either the ALICE or LHCb experiment.

The TRAIN simulation, which takes into account these particularities of the filling scheme and the partial separation in ALICE and LHCb, agrees well with the measured separation in both planes, and all expected features are well reconstructed.

\section{Conclusions}
Emittance scans were used in LHC proton physics operation since 2015 as a complementary, independent tool for measuring the bunch-by-bunch transverse emittances and closed orbit differences in collisions. One scan pair (horizontal and vertical plane) has been done regularly at the start of collisions and before programmed dumps at the CMS experiment. With the operational parameters established, one scan consists of 7 points at different transverse separation, and the full measurement takes less than $5 \ut{min}$.

With the beam and machine parameters used since 2016, the present approach of measuring the bunch length with a full-width half-maximum (FWHM) algorithm and then assuming all planes to be Gaussian lead to systematic errors of up to $\app 30\ut{\%}$ in the crossing plane. To mitigate this, we propose an analysis to take into account the measured longitudinal profiles. This new approach has been validated in simulations and machine studies. Alternatively, if a longitudinal bunch profile measurement is not available, numeric simulations can be used to derive correction constants for the FWHM analysis. For reproducible machine conditions, such a correction can reduce the uncertainty due to the longitudinal profile down to $\app 5\ut{\%}$.

The transverse emittances measured through emittance scans have been compared to measurements from the synchrotron light monitors and the expected values from the ATLAS and CMS luminosities. An agreement better than $20\%$ in convoluted emittance was found over the full run periods of 2015 and 2016. At a plane-by-plane, bunch-by-bunch level, the agreement between the synchrotron light monitors and the emittance scans was found to be better than $10\%$. The uncertainty is dominated by systematic effects which affect the absolute emittance scales for all bunches; the individual, relative bunch-by-bunch measurement is more accurate.

The bunch-by-bunch closed orbit differences observed during the scans are fully explained by beam-beam long-range and PACMAN effects. All bunch-by-bunch features and structures agree with simulation results from the TRAIN code. It is to be noted that over the course of this study, it was found that for the 2017 ``ATS'' LHC optic \cite{ats} the long-range beam-beam encounters beyond the drift space and inner triplets around the IPs have a significant effect, in particular in the horizontal plane. As a consequence, the TRAIN code was improved to include these additional encounters. This highlights the importance of the emittance scan data for benchmarking and improving the simulation code which can then be applied to different scenarios and machines.

Over the course of 2017 LHC operation, the data from emittance scans could also be used by the CMS experiment to continuously check the stability of their individual luminosity monitors over the course of the LHC proton physics runs \cite{olena}. Following these results, the ATLAS collaboration requested emittance scans to be commissioned at their experiment as well for 2018. Since the beam sizes at ATLAS and CMS are the same but the beams cross in different planes, comparing the scans at the two experiments will allow a direct measurement of the impact of the longitudinal distribution in the crossing planes.

\section{Acknowledgements}
We would like to thank the CMS collaboration for agreeing on performing these scans on a regular basis in their experiment, at a small price paid in luminosity. In particular our thank goes to the CMS luminosity team, A.~Dabrowski, M.~Guthoff, O.~Karacheban, P.~Lujan, C.~Palmer, D.~P.~Stickland, and P.~Tsrunchev for many fruitful discussions, and for providing the online luminosity data and error information crucial to this analysis.

W.~Herr, T.~Persson and W.~Kozanecki provided crucial support and information on beam-beam effects, optics simulation codes, and the systematic errors of van-der-Meer like scans respectively. Also, the assistance of the beam instrumentation and RF experts, in particular G.~Trad, M.~Palm and J.~Esteban Muller, is warmly acknowledged.

For many fruitful discussions, suggestions and simulation expertise, we thank the LHC luminosity and beam-beam studies team: F.~Antoniou, I.~Efthymiopoulos, G.~Iadarola, N.~Karastathis, S.~Papadopoulou, and D.~Pellegrini.

We also thank the LHC shift crews for carrying out the emittance scans throughout the LHC proton physics runs of 2015, 2016 and 2017.

This work was supported by the Swiss State Secretariat for Education, Research and Innovation SERI.

\null


\begin{thebibliography}{0}%
\makeatletter
\providecommand \@ifxundefined [1]{%
 \@ifx{#1\undefined}
}%
\providecommand \@ifnum [1]{%
 \ifnum #1\expandafter \@firstoftwo
 \else \expandafter \@secondoftwo
 \fi
}%
\providecommand \@ifx [1]{%
 \ifx #1\expandafter \@firstoftwo
 \else \expandafter \@secondoftwo
 \fi
}%
\providecommand \natexlab [1]{#1}%
\providecommand \enquote  [1]{``#1''}%
\providecommand \bibnamefont  [1]{#1}%
\providecommand \bibfnamefont [1]{#1}%
\providecommand \citenamefont [1]{#1}%
\providecommand \href@noop [0]{\@secondoftwo}%
\providecommand \href [0]{\begingroup \@sanitize@url \@href}%
\providecommand \@href[1]{\@@startlink{#1}\@@href}%
\providecommand \@@href[1]{\endgroup#1\@@endlink}%
\providecommand \@sanitize@url [0]{\catcode `\\12\catcode `\$12\catcode
  `\&12\catcode `\#12\catcode `\^12\catcode `\_12\catcode `\%12\relax}%
\providecommand \@@startlink[1]{}%
\providecommand \@@endlink[0]{}%
\providecommand \url  [0]{\begingroup\@sanitize@url \@url }%
\providecommand \@url [1]{\endgroup\@href {#1}{\urlprefix }}%
\providecommand \urlprefix  [0]{URL }%
\providecommand \Eprint [0]{\href }%
\providecommand \doibase [0]{http://dx.doi.org/}%
\providecommand \selectlanguage [0]{\@gobble}%
\providecommand \bibinfo  [0]{\@secondoftwo}%
\providecommand \bibfield  [0]{\@secondoftwo}%
\providecommand \translation [1]{[#1]}%
\providecommand \BibitemOpen [0]{}%
\providecommand \bibitemStop [0]{}%
\providecommand \bibitemNoStop [0]{.\EOS\space}%
\providecommand \EOS [0]{\spacefactor3000\relax}%
\providecommand \BibitemShut  [1]{\csname bibitem#1\endcsname}%
\let\auto@bib@innerbib\@empty
\end{thebibliography}%


\begin{thebibliography}{99} 
\bibitem{vdm}S.~van der Meer, ``Calibration of the effective beam height in the ISR'', ISR-PO-68-31, CERN, Geneva, Switzerland, 1968.
\bibitem{designreport} O.~Bruning \textit{et al.} (editors), ``LHC Design Report Vol 1: The Main LHC Ring'',  CERN, Geneva, Switzerland, 2004.
\bibitem{cmslumi} The CMS Collaboration, ``CMS luminosity measurement for the 2017 data-taking period at $\sqrt{s} = 13~\mathrm{TeV}$'', CERN, Geneva, Switzerland, 2018.
\bibitem{beambeam} W.~Herr, ``Features and implications of different LHC crossing schemes'',  LHC Project Report 628, CERN, Geneva, 2003.
\bibitem{pacmanssc} P.~L.~Morton and J.~F.~Schonfeld, ``Long Range Beam-Beam Effects'', Proceedings of the Accelerator Physics Issues For A Superconducting Super Collider Conference, Ann Arbor, USA, 1983.
\bibitem{pacmanssc2} N.~K.~Mahale and S.~Ohnuma, ``Beam-beam interaction and pacman effects in the SSC with momentum oscillation'', Particle Accelerators, vol.~27, pp.~175-180, 1990.
\bibitem{pacmantev} V.~Shiltsev and A.~Valishev, ``Beam–Beam Effects'', Accelerator Physics at the Tevatron Collider, ISBN 978-1-4939-0884-4, pp.~411-437, Springer, USA, 2014.
\bibitem{pacman} W.~Herr, ``Effects of PACMAN bunches in the LHC'', LHC Project Report 39, CERN, Geneva, 1996.
\bibitem{pacman2} W.~Herr \textit{et al.}, ``Long-range beam-beam effects in the LHC'', Proceedings of the ICFA Mini-Workshop on Beam-Beam Effects in Hadron Colliders, CERN, Geneva, Switzerland, 2013.
\bibitem{tatianapacman}T.~Pieloni \textit{et al.}, ``Beam-beam effects long-range and head-on'', Proceedings of the 6th Evian Workshop on LHC beam operation, Evian, 2015.
\bibitem{lumi}W.~Herr, ``Particle Colliders and Concept of Luminosity'', at CERN Accelerator School, Granada, Spain, 2012.
\bibitem{tobiasoptics} T.~Persson \emph{et al.}, ``LHC optics commissioning: A journey towards 1\% optics control'', Phys. Rev. ST Accel. Beams \textbf{20}, 061002, 2017.
\bibitem{evian16} M.~Hostettler \emph{et al.}, ``How well do we know our beams?'', Proceedings of the 7th Evian Workshop on LHC beam operation, Evian, 2016.
\bibitem{rfnoise} P.~Baudrenghien \textit{et al.}, ``Longitudinal Emittance Blowup in the Large Hadron Collider'', Nuclear Instruments and Methods in Physics Research Sec. A, vol.~726, pp.~181-190, 2013.
\bibitem{stefania} S.~Papadopoulou \textit{et al.}, ``Modelling and Measurements of Bunch Profiles at the LHC'', Proceedings of the 8th International Particle Accelerator Conference, paper TUPVA044, Kopenhagen, Denmark, 2017.
\bibitem{georgebb} G.~Trad, ``Accounting for longitudinal distributions for emittance calculation from luminosity scans'', Presented at the LHC Beam-Beam and Luminosity Working Group Meeting, 2016.
\bibitem{paper2015}M.~Hostettler \textit{et al.}, ``Beam Size Estimation from Luminosity Scans at the LHC during 2015 Proton Physics Operation'', Proceedings of the 7th International Particle Accelerator Conference,  paper MOPMR025, Busan, Korea, 2016.
\bibitem{juanthesis} J.~Esteban Mueller, ``Longitudinal Intensity Effects in the CERN Large Hadron Collider'', PhD thesis, CERN, Geneva, Switzerland, 2016.
\bibitem{jorgbb} J.~Wenninger \emph{et al.}, ``Observation of Beam-beam Deflections with LHC Orbit Data'', CERN-ACC-NOTE-2013-0006, CERN, Geneva, Switzerland, 2013.
\bibitem{cmsbb} A.~Babaev, ``Beam-dynamic effects at the CMS BRIL van der Meer scans'', JINST \textbf{13} C03028, 2018.
\bibitem{bsrt} G.~Trad \emph{et al.}, ``Performance of the upgraded synchrotron radiation diagnostics at the LHC'',  Proceedings of the 7th International Particle Accelerator Conference,  paper MOPMR030, Busan, Korea, 2016.
\bibitem{bcms} H.~Damerau \textit{et al.}, ``RF Manipulations for Higher Brightness LHC-Type Beams'', Proceedings of the 4th International Particle Accelerator Conference, paper WEPEA044, Shanghai, China, 2013.
\bibitem{imbalance} M.~Hostettler \textit{et al.}, ``Impact of the Crossing Angle on Luminosity Asymmetries at the LHC in 2016 Proton Physics Operation'', Proceedings of the 8th International Particle Accelerator Conference, paper TUPVA005, Kopenhagen, Denmark, 2017.
\bibitem{fanouriamodel} F.~Antoniou, ``Can we predict Luminosity?'', Proceedings of the 7th Evian Workshop on LHC beam operation, Evian, 2016. 
\bibitem{train} H.~Grote and W.~Herr, ``Self-consistent Orbits for beam-beam interactions in the LHC'', Proceedings of the 7th European Particle Accelerator Conference, Vienna, Austria, 2000.
\bibitem{arekthesis} A.~A.~Gorzawski, ``Luminosity control and beam orbit stability with beta star leveling at LHC and HL-LHC'', PhD thesis, Ecole Polytechnique Federale de Lausanne, Switzerland, 2016.
\bibitem{vdmbb} W.~Herr \textit{et al.}, ``Observations of Beam-beam Effects at High Intensities in the LHC'', Proceedings of the 2nd International Particle Accelerator Conference, paper WEODA01, San Sebastian, Spain, 2011.
\bibitem{ats}S.~Fartoukh, ``Achromatic telescopic squeezing scheme and application to the LHC and its luminosity upgrade'', Phys. Rev. ST Accel. Beams \textbf{16}, 111002, 2013.
\bibitem{arekipac}A.~A.~Gorzawski \textit{et al.}, ``Long-Range Beam-Beam Orbit Effects in LHC, Simulations and Observations From Machine Operation in 2016'', Proceedings of the 8th International Particle Accelerator Conference, paper THPAB042, Kopenhagen, Denmark, 2017.
\bibitem{olena} O.~Karcheban \textit{et al.}, ``Emittance scans for CMS luminosity calibration'', Presented at the LHCC Poster Session, CERN, Geneva, Switzerland, 2018.

\end{thebibliography}
\end{document}